\documentclass[12pt]{article}

\begin{document}

\def\CA{{\cal A}}
\def\CB{{\cal B}}
\def\CC{{\cal C}}
\def\CD{{\cal D}}
\def\CE{{\cal E}}
\def\CF{{\cal F}}
\def\CG{{\cal G}}
\def\CH{{\cal H}}
\def\CI{{\cal I}}
\def\CJ{{\cal J}}
\def\CK{{\cal K}}
\def\CL{{\cal L}}
\def\CM{{\cal M}}
\def\CN{{\cal N}}
\def\CO{{\cal O}}
\def\CP{{\cal P}}
\def\CQ{{\cal Q}}
\def\CR{{\cal R}}
\def\CS{{\cal S}}
\def\CT{{\cal T}}
\def\CU{{\cal U}}
\def\CV{{\cal V}}
\def\CW{{\cal W}}
\def\CX{{\cal X}}
\def\CY{{\cal Y}}
\def\CZ{{\cal Z}}

\newcommand{\btzm}[0]{BTZ$_{\rm M}$}
\newcommand{\todo}[1]{{\em \small {#1}}\marginpar{$\Longleftarrow$}}
\newcommand{\labell}[1]{\label{#1}\qquad_{#1}} 
\newcommand{\ads}[1]{{\rm AdS}_{#1}}
\newcommand{\SL}[0]{{\rm SL}(2,\IR)}
\newcommand{\cosm}[0]{R}
\newcommand{\tL}[0]{\bar{L}}
\newcommand{\hdim}[0]{\bar{h}}
\newcommand{\bw}[0]{\bar{w}}
\newcommand{\bz}[0]{\bar{z}}
\newcommand{\lp}{\lambda_+}
\newcommand{\bx}{ {\bf x}}
\newcommand{\bk}{{\bf k}}
\newcommand{\bb}{{\bf b}}
\newcommand{\BB}{{\bf B}}
\newcommand{\tp}{\tilde{\phi}}
\hyphenation{Min-kow-ski}

\def\ie{{\it i.e.}}
\def\eg{{\it e.g.}}
\def\cf{{\it c.f.}}
\def\etal{{\it et.al.}}
\def\etc{{\it etc.}}

\def\apr{\alpha'}
\def\str{{str}}
\def\lstr{\ell_\str}
\def\gstr{g_\str}
\def\Mstr{M_\str}
\def\lpl{\ell_{pl}}
\def\Mpl{M_{pl}}
\def\varep{\varepsilon}
\def\del{\nabla}
\def\grad{\nabla}
\def\tr{\hbox{tr}}
\def\perp{\bot}
\def\half{\frac{1}{2}}
\def\p{\partial}

\let\a=\alpha\let\b=\beta\let\d=\delta
\let\e=\epsilon\let\f=\phi\let\g=\gamma
\let\h=\eta\let\th=\theta\let\k=\kappa\let\l=\lambda
\let\m=\mu\let\n=\nu\let\p=\pi\let\r=\rho
\let\s=\sigma\let\t=\tau\let\u=\upsilon
\let\w=\omega\let\x=\xi\let\y=\psi
\let\z=\zeta\let\G=\Gamma\let\P=\Pi\let\S=\Sigma
\let\Th=\Theta\let\D=\Delta\let\F=\Phi
\let\vt=\vartheta \let \L=\Lambda
\newcommand{\hmu}{\hat{\mu}}
\newcommand{\hnu}{\hat{\nu}}
\newcommand{\hrho}{\hat{\rho}}
\newcommand{\hgam}{\hat{\gamma}}
\newcommand{\hpsi}{\hat{\psi}}
\newcommand{\hPsi}{\hat{\Psi}}
\newcommand{\heps}{\hat{\epsilon}}
\newcommand{\hxi}{\hat{\xi}}
\newcommand{\hm}{\hat{m}}
\newcommand{\hn}{\hat{n}}
\newcommand{\ha}{\hat{a}}
\newcommand{\hb}{\hat{b}}
\newcommand{\nn}{\nonumber}
\newcommand{\un}{\underline}
\newcommand{\be}{\begin{equation}}
\newcommand{\ee}{\end{equation}}
\newcommand{\bea}{\begin{eqnarray}}
\newcommand{\eea}{\end{eqnarray}}
\newcommand{\eps}{\epsilon}
\newcommand{\kap}{\kappa}
\newcommand{\Pp}{{\cal P}_+}
\newcommand{\Pm}{{\cal P}_-}
\newcommand{\Ppm}{{\cal P}_\pm}
\newcommand{\dk}{\delta_\kappa}
\newcommand{\T}{\Theta}
\newcommand{\ksym}{$\kappa$-symmetry }
\renewcommand{\paragraph}[1]{
\vspace{.8mm}\par\noindent {\sl #1}\\
\vspace{0.2mm} }
\def\SLK{\hbox{\ooalign{$\displaystyle \Lambda_K$\cr$\hspace{2pt}/$}}}
\newcommand{\bos}{^{\bf B}}
\newcommand{\ferm}{^{\bf F}}
\newcommand{\rep}{{\cal G}}
\newcommand{\eqn}[1]{(\ref{#1})}
\newcommand{\ft}[2]{{\textstyle\frac{#1}{#2}}}
\newcommand{\Poin}{Poincar\'e}
\renewcommand{\u}[1]{{\bar{#1}}}
\newcommand{\ba}{\left(\begin{array}}
\newcommand{\ea}{\end{array}\right)}
\newcommand{\PL}{{\cal P}_L{}}
\newcommand{\PR}{{\cal P}_R{}}
\newcommand{\M}{{\cal M}}
\newcommand{\R}{{\cal R}}
\newcommand{\A}{{\cal A}}
\newcommand{\ta}{\dot{\tilde \a}}
\newcommand{\tb}{\dot{\tilde \b}}
\newcommand{\tc}{\dot{\tilde \c}}
\newcommand{\td}{\dot{\tilde \d}}
\newcommand{\dta}{\dot{\tilde \a}}
\newcommand{\dtb}{\dot{\tilde \b}}
\newcommand{\dtc}{\dot{\tilde \c}}
\newcommand{\dtd}{\dot{\tilde \d}}
\renewcommand{\L}{{\cal L}}
\newcommand{\Tf}{\Theta_f}
\newcommand{\tf}{\theta_f}
\def\da{{\dot\alpha}}
\def\ddt{{\partial_\tau}}
\def\Hgp{{\bf H}}             
\def\Ggp{{\bf G}}
\def\Kgp{{\bf K}}
\def\dmin{{\partial_-}}
\def\dpl{{\partial_+}}
\newcommand{\Ka}{{\bf K}} 
\newcommand{\Ha}{{\bf H}}
\newcommand{\Ga}{{\bf T}}
\def\H{{\bf H}}
\def\K{{\bf K}}
\def\G{{\bf G}}
\def\Hl{{\bf H}}
\def\Kl{{\bf K}}
\def\Gl{{\bf G}}
\def\Tl{{\bf T}}
\def\gg{{\cal G}}
\def\lie{{\cal L}}
\def\mb{{\bar M}}
\def\su{{{\rm SU}(2,2|4)}}

\def\hn{\hat{n}}
\newsavebox{\uuunit}
\sbox{\uuunit}
{\setlength{\unitlength}{0.825em}
 \begin{picture}(0.6,0.7)
\thinlines
\put(0,0){\line(1,0){0.5}}
\put(0.15,0){\line(0,1){0.7}}
 \put(0.35,0){\line(0,1){0.8}}
\multiput(0.3,0.8)(-0.04,-0.02){12}{\rule{0.5pt}{0.5pt}}
\end {picture}}
\newcommand {\unity}{\mathord{\!\usebox{\uuunit}}}
\renewcommand{\theequation}{\thesection.\arabic{equation}}


\newpage
\renewcommand{\thepage}{\arabic{page}}
\setcounter{page}{1}

\rightline{CALT68-2290, CITUSC/00-39}
\rightline{HUTP-00/A031, UPR-897T}
\rightline{hep-th/0007211}
\vskip 1cm
\centerline{\Large \bf Four Dimensional Conformal Supergravity From AdS Space}
\vskip 1cm

\renewcommand{\thefootnote}{\fnsymbol{footnote}}
\centerline{{\bf Vijay
Balasubramanian${}^{1}$\footnote{vijayb@pauli.harvard.edu.
Address  from September 2000: 
\\ \hspace*{0.2in} 
David Rittenhouse Laboratories,
University of Pennsylvania, Philadelphia, Pennsylvania 19104.}, 
Eric Gimon${}^{2}$\footnote{egimon@theory.caltech.edu.},
Djordje Minic${}^{3}$\footnote{minic@physics.usc.edu.},
 and
Joachim Rahmfeld${}^{2}$\footnote{rahmfeld@theory.caltech.edu.}}}
\vskip .5cm
\centerline{${}^1$\it Jefferson Laboratory of Physics, Harvard
University,}
\centerline{\it Cambridge, MA 02138, USA.}
\vskip .5cm
\centerline{${}^2$ \it  CIT-USC Center for Theoretical Physics,}
\centerline{\it California Institute of Technology,  Pasadena, CA 91125}

\vskip .5cm
\centerline{${}^3$ \it  CIT-USC Center for Theoretical Physics,}
\centerline{\it Department of Physics and Astronomy,}
\centerline{\it University of Southern California,  Los Angeles, CA
90089-0484.}

\setcounter{footnote}{0}
\renewcommand{\thefootnote}{\arabic{footnote}}

\begin{abstract}
Exploring the role of conformal theories of gravity in string theory,
we show that the minimal ($N=2$) gauged supergravities in five
dimensions induce the multiplets and transformations of $N=1$ four
dimensional conformal supergravity on the spacetime boundary.  $N=1$
Poincar\'e supergravity can be induced by explicitly breaking the
conformal invariance via a radial cutoff in the 5d space.  The AdS/CFT
correspondence relates the maximal gauged supergravity in five
dimensions to $N=4$ super Yang-Mills on the $4d$ spacetime boundary.
In this context we show that the conformal anomaly of the gauge theory
induces conformal gravity on the boundary of the space and that this
theory, via the renormalization group, encapsulates the gravitational
dynamics of the skin of asymptotically AdS spacetimes.  Our results
have several applications to the AdS/CFT correspondence and the
Randall-Sundrum scenario.
\end{abstract}

\section{Introduction}
\label{intro}

The classic methods of Kaluza and Klein \cite{KK} are the conventional
tools in supergravity for generating the dynamics of a lower
dimensional space from a higher dimensional world with a compact
factor.  The light fields in the lower dimensions arise from
fluctuations that solve massless wave equations on the internal space,
and the symmetries governing their dynamics are derived by appropriate
restriction from the higher dimensions.  Recently, string theorists
and phenomenologists have studied the physics of worlds that exist on
branes or submanifolds embedded in a higher dimensional space.  In the
AdS/CFT \cite{juanads,gkpw} and Randall-Sundrum \cite{rs} contexts,
the relevant 4-surface lies near or at the boundary of a five
dimensional space which asymptotically has a negative cosmological
constant.  Such asymptotically anti-de Sitter (AdS) spaces arise
naturally as solutions to 5d gauged supergravities \cite{murat,PPN},
or as the near horizon limits of string compactifications containing
3-branes \cite{DuffLuHS}.  One purpose of this article is to show in
detail how $N=1$ supergravity is induced on such surfaces when the
bulk theory enjoys $N=2$ supersymmetry.  When the bulk is non-compact,
the $N=1$ theory is conformal.\footnote{See~\cite{csgrev,liutsey} for
reviews of conformal gravity.}   Poincar\'e supergravity can be
regained by cutting off the bulk space to explicitly break conformality. 

The AdS/CFT correspondence states that the classical action for an
asymptotically AdS space, regulated by boundary counterterms (see,
e.g.,~\cite{HenSken,adsstress}) and treated as a functional of boundary data,
is equal to the effective action for an $N=4$ super Yang-Mills (SYM)
theory.  This theory has a conformal anomaly~\cite{riegert} which
reconstructs the action of $N=4$ conformal supergravity in four
dimensions.  By the AdS/CFT correspondence this must be related to a
log divergent term of the spacetime action \cite{HenSken,ejm}.  This
implies that the asymptotically AdS solutions to $N=8$ gauged
supergravity induce $N=4$ conformal gravity on the spacetime boundary.
Turning this analysis around, the complete gravitational dynamics of
the skin of the spacetime is reproduced holographically
\cite{holog,susswitt,amanda} by the conformal anomaly of the dual
Yang-Mills theory, thus lending further support to the holographic RG
setup \cite{adsrg,adsrgothers,hermanrg,klebwitt,adsrgsolns,vatche}.

Conformal gravity remains one of the few classical theories of gravity that
has not been integrated into the framework of string theory.  The paper
concludes with speculations about the role of conformal gravity, and
discusses some applications of our results in the AdS/CFT and
Randall-Sundrum contexts.

\setcounter{equation}{0}
\section{One supersymmetry from two}
\label{onesusy}

The pure $N=2$, $d=5$ gauged supergravity~\cite{murat} admits solutions
that asymptotically have constant negative curvature.  We seek the residual
symmetries induced on the boundary of such spaces by the bulk theory.  It
will transpire that the boundary fields transform in multiplets of the
$N=1$, $d=4$, conformal supergravity (listed in~\cite{csgrev}, sec. 2.2).
Maximally supersymmetric gauged supergravities in $d=3,6,7$ were related to
conformal supergravities in $d=2,5,6$ in \cite{tanii}.
We will first present the fields and symmetries of the pure $N=2$ gauged
supergravity in five dimensions, and then argue that the boundary values
have the correct multiplicities to form the gravity multiplet of 4d $N=1$
conformal supergravity.  Finally, we show that the radial
diffeomorphisms and supersymmetries of the bulk induce the symmetries of
conformal supergravity on the four dimensional boundary.

\vskip 0.15in
\paragraph{$\bullet$ $N=2$ gravity in five dimensions} 
The gravity multiplet of the minimal gauged supergravity in five
dimensions consists of the f\"unfbein $\hat e_{\hat \mu}^{\hat a}$,
two gravitinos ${\hat \psi}_{\hat\rho i}$, and a gauge field $\hat
A_{\hat \mu}$, where $i=1,2$. The gravitinos are related by the
pseudo-symplectic Majorana condition. A $U(1)$ subgroup of the $SU(2)$
automorphism group of the $N=2$ algebra is gauged, and the field $\hat
A_{\hat \mu}$ serves as the corresponding gauge field. The Lagrangian
of the theory is then given (up to four-fermi terms) by \cite{murat}
\begin{eqnarray}
\hat e^{-1}\hat {\cal L}_5&=&-\frac{1}{2} \hat R-\frac{1}{2}
\hat{{{\bar \psi}}}_{\hat\mu}^i\hat \gamma^{\hat\mu\hat\nu\hat\rho}
\hat {\cal D}_{\hat\rho}
{\hat  \psi}_{\hat\rho i}-\frac{3R^2}{32}\hat F_{\hat\mu\hat\nu}\hat
F^{\hat\mu\hat\nu}-\frac{3i}{4R}\hat {\bar \psi}_{\hat\mu}^i
\hat \gamma^{\hat\mu\hat\nu}{\hat  \psi}_{\hat\nu}^j \d_{ij}+
\frac{6}{R^2}\, , \nn \\
& & -\frac{3iR}{32}\left( \hat {\bar \psi}_{\hat\mu}^i \hat
\gamma^{\hat \mu\hat \nu\hat \rho\hat \sigma}{\hat  \psi}_{\hat\nu i}
\hat F_{\hat \rho\hat \sigma}+2\hat {\bar \psi}^{\hat\mu i}
{\hat  \psi}^{\hat\nu}_{i}\hat F_{\hmu\hnu}   \right)+\frac{\hat e^{-1}}
{6\sqrt{6}}c \hat \eps^{\hat \mu\hat \nu\hat \rho\hat \sigma\hat \lambda}
\hat F_{\hmu\hnu}\hat F_{\hat \rho\hat \sigma}\hat A_{\hat \lambda}\, ,
\label{gsgAction}
\end{eqnarray}
where $c$ is a constant.  Typically, $N=2$ theories that are obtained in 5d
by compactification of M-theory on a Calabi-Yau threefold include
additional hypermultiplets containing the moduli of the compact space,
including the 5d dilaton.  Likewise, the multiplets of the maximal $N=8$
gauged supergravity can be decomposed in terms of an $N=2$ gravity
multiplet, along with some hypermultiplets and vector multiplets.  All of
these multiplets may be studied by methods similar to those used below to
study the $N=2$ gravity multiplet.

The normalization is chosen so that the vacuum is $AdS_5$
space with radius $R$:
\begin{equation}
ds^2 =
\frac{R^2}{r^2} (dx^{\mu} dx^{\nu} \eta_{\mu \nu} + dr^2)\, .
\label{StandardMetric}
\end{equation}
Here, $ \mu = 0, 1,2,3$, and $r$ is the radial direction with the
spacetime boundary at $r=0$.  For reference, in the notation
of~\cite{murat} our conventions are $g=\frac{3}{4}, \
P_0=\frac{4}{R}\sqrt{\frac{2}{3}}, \
V_1=1, \ h_1=\frac{R}{2}\sqrt{\frac{3}{2}}$ and $h^1=1/h_1$.
Hats denote 5d objects, so that $\hat \mu=0,1,2,3,4$ etc., and
the signature is $(-++++)$. Also,
the first part of the Latin alphabet is
reserved for tangent space indices. The gauge covariant derivative
is
\begin{equation}
(\hat {\cal D}_{\hat \mu}(\hat \omega) \hat \psi_{\hat \nu})^i=
\hat {D}_{\hat \mu}(\hat \omega) \hat \psi_{\hat \nu}^i+\frac{4}{3}
\hat A_{\hat \mu}\d^{ij}\hat \psi_{\hat \nu j}\,
\end{equation}
in terms of $\hat {D}_{\hat \mu}(\hat \omega)$, the standard covariant
derivative
\begin{equation}
\hat {D}_{\hat \mu}(\hat \omega) \hat \psi_{\hat \nu}^i=
\partial_{\hat \mu}\hat \psi_{\hat \nu}^i+\frac{1}{4}\hat
\omega_{\hat \mu}^{\hat a \hat b}\psi_{\hat \nu}^i\, .
\end{equation}
Finally, $i,j$ indices are raised with the epsilon symbol:
$
\psi^i=\eps^{ij} \psi_j, \ \eps^{12}=1\, .
$

The local supersymmetry variations are
\begin{eqnarray}
\hat \d \hat e_{\hat \mu}^{\hat a}&=&\frac{1}{2}\hat {\bar \epsilon}^i
   \hat \gamma^{\hat a} \hat \psi_{\mu i} \label{GaugedSugraE}  \\
\hat \d  \hat \psi_{\hat \mu}^i&=& \hat {\cal D}_{\hat \mu}(\hat {\tilde
\omega})
\hat \e_i+\frac{i R}{8}\left(\hat \gamma_{\hmu}{}^{\hnu\hrho}-4
\d_{\hmu}^{\hnu}\hat \gamma^{\hrho} \right) \hat {\tilde
F}_{\hnu\hrho}\hat \e_i
+\frac{i}{2 R}\hat \gamma_{\hmu}\d^{ij}\hat \e_j\label{GaugedSugraPsi}
\\
\hat \d \hat A_{\hat \mu}&=& \frac{i}{R}
{\hat {\bar \psi}}_{\hat\mu}^i\e_i\label{GaugedSugraA}
\, .
\end{eqnarray}
Here,
\begin{eqnarray}
\hat {\tilde \omega}_{\hmu \ha \hb}&=&\hat {\omega}_{\hmu \ha
\hb}-\frac{1}{4}
\left(
{\hat {\bar \psi}}^i_{\hb} \hgam_{\hmu} \hat \psi_{\ha i}+2
{\hat {\bar \psi}}^i_{\hmu}\hgam_{[\hb} \hat \psi_{\ha] i}\right) \nn \\
\hat {\tilde F}_{\hmu\hnu}&=& \hat {F}_{\hmu\hnu}+\frac{i\sqrt{6}}{4}
{\hat {\bar \psi}}^i_{[\hmu}{\hat \psi}_{\hnu] i}
\end{eqnarray}
In order to study the boundary limit of these supersymmetry
variations, we will also require the radial coordinate
transformations, parametrized by $\xi^r$:
\begin{eqnarray}
\hat \d_{\hxi} \hat e_{\hat \mu}^{\hat a}&=&
\hxi^r\partial_r  \hat e_{\hat \mu}^{\hat a} +
\partial_{\hmu} \hxi^r  \hat e_{r}^{\hat a} \label{ctrans1} \\
\hat \d_{\hxi}  \hat \psi_{\hat \mu}^i&=&
\hxi^r\partial_r  \hat \psi_{\hat \mu}^i +
\partial_{\hmu} \hxi^r   \hat \psi_{r}^i\label{dilatationsPsi}
\label{ctrans2} \\
\hat \d_{\xi} \hat A_{\hat \mu}&=&
\hxi^r\partial_r \hat A_{\hat \mu} +
\partial_{\hmu} \hxi^r   \hat A_{r}
\, .
\label{ctrans3}
\end{eqnarray}
In the boundary limit ($r \rightarrow 0$), these equations will act as
conformal transformations, which, together with the induced
supersymmetries, will reproduce the symmetries of four dimensional
conformal supergravity.

\vskip 0.15in
\paragraph{$\bullet$ Conformal supergravity multiplet}
To begin we must identify what we mean by the ``boundary degrees of
freedom'' which enjoy the symmetries of 4d conformal gravity.  The
vacuum solution to the equations of motion of 5d gauged supergravity
is $\ads{5}$ space, which is non-compact and only has a boundary in
the conformal sense.  In perturbation around this background,
solutions to the equations of motion either vanish or diverge at
infinity.  (This is the familiar split into normalizable and
non-normalizable modes in the AdS/CFT correspondence~\cite{bkl}.)
Here we will argue that a similar split holds for the fully non-linear
equations of motion.  The boundary fields are identified as the finite
residue that remains after removing the scaling divergence of
non-normalizable bulk fields.  Notably, the non-chiral fermion of the
bulk theory loses half its components in the process and becomes
chiral. In this way, the boundary values of bulk fields realize the
gravity multiplet of $N=1$, $d=4$ conformal supergravity.

To proceed, set $\hat{\psi} = 0$ and $\hat{A} = 0 $.  Then Fefferman
and Graham~\cite{fefgraham} have shown that near the boundary (at
$r=0$), a general solution to the equations of motion can be written
as
\begin{equation}
ds^2 = {R^2 \over r^2} (dx^\mu \, dx^\nu \, g_{\mu\nu} + dr^2)
\label{gensoln}
\end{equation}
where $g_{\mu\nu} = g^0_{\mu\nu}(x) + O(r^2)$.  In the language
of~\cite{bkl}, the $O(1/r^2)$ piece of the metric is the
non-normalizable mode whose boundary value determines a conformal
class of boundary metrics: the divergence at $r=0$ is removed by
multiplying the metric by any function scaling as $r^2$ as $r
\rightarrow 0$, giving $\Omega(x)^2 \, g^0_{\mu\nu}(x)$ as the
boundary value of the metric.  Equivalently, different rates of
approach to $r =0$ at different boundary positions $x$ yields a
conformal factor $\Omega(x)^2$.

Hence, the radial diffeomorphisms (\ref{ctrans1}) become conformal
transformations of the boundary metric as $r \rightarrow 0$ provided
\begin{equation}
\hxi^r\equiv r \lambda_D\, .
\label{dilat}
\end{equation}
Such diffeomorphisms ``warp'' surfaces homeomorphic to the AdS
boundary in the radial direction, producing different conformal
factors in the limiting procedure that yields the boundary
metric~\cite{deform}.  The $1/r^2$ radial dependence of the leading
term in the bulk metric determines that the boundary metric
$g^0_{\mu\nu}(x)$ has a conformal weight of $2$.  Similarly, we will
see that the radial dependence of other bulk fields also determines
their boundary conformal weight.

We choose $g^0_{\mu\nu}$ as the representative of the boundary
conformal class, and partially fix the local symmetries.
Following existing examples constructed by Nishimura and Tanii 
in $d=4,6,7$~\cite{tanii}, set:
\begin{eqnarray}
\hat e^a_\mu(r,x)&=&\frac{R}{r} \,e^a_\mu(x) +O(1) \, ,
\qquad \hat e^a_r=e^{\underline
r}_\mu=0 \, ,
\qquad
\hat e^{\underline r}_r=\frac{R}{r} \, , \label{fixVielbein} \\
\hpsi_r^i&=& 0 \, ,\\
\hat A_r&=& 0 \, .\nn
\end{eqnarray}
($\underline r$ is the radial tangent space index.)  We will examine
the linearized bulk equations of motion for $\hat A_\mu$ and $\psi$ in
the above background and gauge, and then argue that the nonlinear
couplings between these fields and the metric do not modify the
asymptotic scaling of solutions.

First examine the $r$-dependence of $\hat A_\mu$.  Asymptotically,
solutions to the equation of motion,
\begin{equation}
\frac{1}{\sqrt{-\hat g}} \partial_{\hat \mu} (\hat g^{\hmu \hnu}
\partial_{\hnu}
\hat A_{\rho}) =0 \, ,
\end{equation}
are found by choosing $\hat{A}_\mu$ to be independent of
the radial direction,
\begin{equation}
\hat A_\mu=A_\mu\, ,
\end{equation}
and solving the resulting four-dimensional equation of motion.  The
radial diffeomorphisms (\ref{ctrans2}) with a dilatation parameter
$\lambda_D$ as in (\ref{dilat}) leave $A_\mu$ invariant.  So we find
that the gauge field has zero weight under the boundary conformal
transformations induced by radial diffeomorphisms.

Next we turn to the gravitino.  After using (\ref{fixVielbein}) the
four- and five-dimensional spin connections become
\begin{eqnarray}
{\hat{\omega}}^{ar} &= &-{ 1 \over r} e^{a}, \nn \\
{\hat{\omega}}^{ab} &= &\omega^{ab}\, ,
\end{eqnarray}
and the covariant derivatives reduce to
\begin{eqnarray}
{\hat{D}}_{\mu} &= &D_{\mu} - {1 \over 2r} \gamma_{\mu} \gamma_{r}\nn
\\
{\hat{D}}_{r} &=& \partial_{r} \label{CovDer}
\end{eqnarray}
To determine the decomposition of the gravitino we use the (linearized)
gravitino equation of motion
\begin{equation}
\hgam^{\hmu\hnu\hrho} {\hat D}_{\hnu}\hpsi_{\hrho i}
-\frac{3i}{2R}\hgam^{\hmu\hnu}\hpsi_{\hnu}^j \d_{ij}=0 \, .
\end{equation}
This equation reduces, after using (\ref{CovDer}), to
\begin{equation}
\gamma^{\mu \rho} (\delta_{ij} \partial_r -
\delta_{ij} {1 \over r} -
\frac{3i}{2r} \epsilon_{ij} \gamma_5) \hpsi_{\rho j}
+ \gamma^{\mu \nu \rho} D_{\nu} \gamma_5 \hpsi_{\rho i} = 0
\end{equation}
where $\gamma_5 \equiv \gamma^{\underline r}$ squares to one.
Note that $\hat\gamma^\mu=\hat e^{\mu}_a \gamma^a
=\frac{r}{R}\gamma^\mu $. These two equations can be diagonalized by
introducing
\begin{equation}
\hPsi_{\rho} \equiv \hpsi_{\rho 1} + i \hpsi_{\rho 2}\, ,
\label{PsiDef1}
\end{equation}
which satisfies
\begin{equation}
\gamma^{\mu \rho} ( \partial_r -
 {1 \over r} -
\frac{3}{2r}  \gamma_5) \hPsi_{\rho }
+ \gamma^{\mu \nu \rho} D_{\nu} \gamma_5 \hPsi_{\rho } = 0.
\label{GravEqn}
\end{equation}
$\Psi$ is then decomposed into a chiral and an anti-chiral
components with respect to $\gamma_5$:
\begin{equation}
\hPsi_{\rho}^{R} \equiv \frac{1}{2}(1-\gamma_5)\hPsi_{\rho}
\end{equation}
\begin{equation}
\hPsi_{\rho}^{L} \equiv \frac{1}{2}(1+\gamma_5)\hPsi_{\rho}\, .
\end{equation}
The dominating solution to (\ref{GravEqn}) is then given by
\begin{equation}
\hPsi_{\rho}^{R}=(\frac{2R}{r})^{\frac{1}{2}}\Psi_{\rho}^{R}\, .
\end{equation}
The radial dependence of  $\hPsi_{\rho}^{R}$ combines with the radial
diffeomorphisms (\ref{ctrans3}) and (\ref{dilat}) to give a conformal
weight of $-1/2$ for the boundary value $\Psi_{\rho}^R$.

It is important not to forget the constrained
components of the gravitino $\hPsi_{\rho}^{L}$, as they enter
the supersymmetry variations.
Let
\begin{equation}
\hPsi_{\rho}^{L}=(2Rr)^{\frac{1}{2}}\Phi_{\rho}^{L}\, ,
\end{equation}
then
$\Phi_{\rho}^L$ satisfies the following equation
\begin{equation}
4 \gamma^{\mu \rho} \Phi_{\rho}^L = \gamma^{\mu \nu \rho} [D_{\nu}
\Psi_{\rho}^R - D_{\rho}
\Psi_{\nu}^R] \, .
\end{equation}
This equation is solved by
\begin{equation}
\Phi_{\rho}^L = {1 \over 3} \gamma^{\nu} [D_{\nu} \Psi_{\rho}^R -
D_{\rho} \Psi_{\nu}^R]
+ { i \over {12}} \gamma_5 \gamma^{\lambda} \epsilon_{\lambda \rho}^{\sigma
\tau} [D_{\sigma} \Psi_{\tau}^R -
D_{\tau} \Psi_{\sigma}^R]\, .
\end{equation}

It only remains to argue that the asymptotic scalings and
resulting four dimensional conformal weights are unchanged when
the non-linear couplings between all the fields are accounted for.
Also, one has to ensure that the ansatz for the 5d metric in
(\ref{fixVielbein}) is consistent in the presence of non-trivial
fields. After all, this was derived in \cite{fefgraham} only for
pure gravity with a cosmological constant.

First we turn to the vierbein. To see that the asymptotic
behavior
\begin{equation}
\hat e_\mu^a(x,r)=\frac{R}{r} e_\mu^a(x)+ {\rm subleading \ terms}\, 
\label{scaling}
\end{equation}
is consistent even with non-trivial fields we study the scalings of 
terms in the bulk action.  If the cosmological constant
dominates in the boundary limit, the Fefferman-Graham analysis which
yielded (\ref{scaling}) will continue to hold.  By definition, the
cosmological constant is $r$-independent. Given the asymptotic form of the
gravitino and gauge field that we have derived, it is easily verified that
the kinetic, interaction and four-fermi terms of the $N=2$ gravity
action (eq. (\ref{gsgAction}) and~\cite{murat}) scale to zero at least as fast
as $\sim r^{2}$ when $r\rightarrow 0$.  Hence their contribution is
subleading and the asymptotic scaling of the vielbein (or metric)  survives
the nonlinear interactions.

The same logic applies to the gravitino and the gauge field.  With the
scaling ansatze we have made for all the fields, the leading behaviour
of the non-normalizable modes of the 5d gravitini is not
altered when the full interaction terms are accounted for.  Neither is the
asymptotic behavior of the gauge field affected. Certainly though, the
non-linear terms in the equations of motion give rise to interactions
between  the various modes.   This is precisely as expected -- conformal
supergravity is not a free theory.

We have shown how to extract the boundary values $(e_\mu^a,\Psi_\mu,
A_\mu)$ of the $N=2$ gauged supergravity multiplet $(\hat
e_{\hmu}^{\ha},\hPsi_{\hmu}^i, \hat A_{\hmu})$ and argued that these fields
transform with specific weights under the induced conformal transformations
of the spacetime boundary.  In particular, although the bulk gravitino is
non-chiral, its boundary value is chiral.  In fact, $(e_\mu^a,\Psi_\mu,
A_\mu)$ is precisely the gravity multiplet of $N=1$ conformal supergravity.
It remains to show that even the supersymmetries of this $N=1$ theory are
induced on the spacetime boundary by the bulk transformations
(\ref{GaugedSugraE})-(\ref{GaugedSugraA}).

\vskip 0.15in
\paragraph{$\bullet$ Conformal supergravity symmetries}
We have already shown that conformal transformations are induced on
the spacetime boundary by bulk radial diffeomorphisms.  To treat the
induced supersymmetries we start with the f\"unfbein. Define
$\heps=\eps_1+i\eps_2$, and decompose the SUSY parameter $\eps$
according to its chirality under $\gamma_5$. Following the analysis of
the 5d gravitino, the two components should have appropriate
scaling factors, so that
\begin{equation}
\heps = (2R)^{\frac{1}{2}}(r^{-1/2} \epsilon^R, r^{1/2} \eta^L)
\end{equation}
with $\gamma_5 \eps^R=-\eps^R,\ \gamma_5\eta^L=\eta^L$.  As the
divergent piece, $\eps$ becomes the 4d supersymmetry parameter,
whereas $\eta$ parametrizes special supersymmetries.  To leading order
in $r$, we derive from (\ref{GaugedSugraE})
\begin{eqnarray}
\delta e^{a}_{\mu} &=&\frac{R}{2r}{\hat{\bar{\epsilon}}}^i
\gamma^a\hat \psi_{\mu i}\nn\\
&=&\frac{R}{2r} {\hat{\bar{\epsilon}}} \gamma^a \hat \Psi_{\mu}\nn \\
&=& {{\bar\epsilon^R}} \gamma^a \Psi_{\mu}^R+O(r)\nn\\
&=&-\frac{1}{2} {{\bar\Psi}_\mu} \gamma^a \epsilon+O(r),
\label{eTraf}
\end{eqnarray}
where we introduced Majorana spinors
\begin{equation}
\chi=\pmatrix{\chi^R \cr \chi^L}, \quad {\rm with} \quad (\chi^R)^*=\eps \chi^L,
\end{equation}
with $\eps=-i\s_2$.
This is nothing but the standard variation
of the vierbein in four dimensions.
Note, that $\eta$ decouples from the transformation of $e_\mu^a$.

Next, we turn to the gauge field. Using (\ref{GaugedSugraA}) gives
\begin{eqnarray}
\delta A_{\mu} &=& 2 i  \left({\bar{\Phi}}_{\mu}^L \epsilon^R
+ {\bar{\Psi}}_{\mu}^R \eta^L\right)\,\nn \\
 &=&  - i  \left({\bar{\Phi}}_{\mu}\gamma_5 \epsilon
- {\bar{\Psi}}_{\mu}\gamma_5 \eta\right) \, ,
\label{ATraf}
\end{eqnarray}
which agrees with the transformation law given in
\cite{csgrev}.

The analysis of the gravitino is more difficult since
(\ref{GaugedSugraPsi}) is rather complicated. Fortunately, many terms
drop out in the boundary limit. First of all, $\hat \d \hPsi \sim
r^{-1/2}$, so the term containing $F$ vanishes, because it scales as
$r^{1/2}$.  Also, the difference between $\hat \omega$ and $\hat
{\tilde \omega}$ disappears, because the bilinears in the gravitino
scale with a higher power of $r$.  One is left with
\begin{equation}
\hat \d  \hat \psi_{\hat \mu}^i\sim  \hat {\cal D}_{\hat \mu}(\hat
{\omega})
\hat \e_i+\frac{i}{2R}\hat \gamma_{\hmu}\d^{ij}\hat
\e_j \, ,
\label{PsiTrafoRes}
\end{equation}
where
\begin{equation}
(\hat {\cal D}_{\hmu} \heps)^{i} = \hat D_{\hmu} \heps^i
+\frac{3}{4} \hat A_{\hmu} \delta^{ij} \heps_{j}
\, .
\end{equation}
With (\ref{PsiDef1}) this translates the chiral
component of the gravitino to
\begin{equation}
\delta \Psi^R = {D}_{\mu} \epsilon^R +\frac{3i}{4}A_\mu \eps^R -\gamma_\mu
\eta^L\, ,
\label{psiTraf}
\end{equation}
implying
\begin{equation}
\delta \Psi^L = {D}_{\mu} \epsilon^L -\frac{3i}{4}A_\mu \eps^L -\gamma_\mu
\eta^R\, .
\label{psiTrafL}
\end{equation}
The last term in this expression has two origins. First, the relation
between five- and four-dimensional covariant derivatives contains an
extra term according to (\ref{CovDer}). Second, it can be shown that
this term gives the same contribution as the last term in
(\ref{PsiTrafoRes}).

We have shown in (\ref{eTraf}), (\ref{psiTraf}) and (\ref{ATraf}) that
the 5d SUSY transformations reduce on the boundary to the residual
transformations:
\begin{eqnarray}
\delta e^{a}_{\mu} &=& -\frac{1}{2} {\bar{\psi}_\mu} \gamma^a \epsilon  \nn \\
\delta \Psi &=& {\cal D}_{\mu} \epsilon -\gamma_\mu \eta\, ,
\label{TrafoFinal}\\
\delta A_{\mu} &=& i \left({\bar{\Psi}}_{\mu} \gamma_5 \eta-
{\bar{\Phi}}_{\mu} \gamma_5\epsilon\right)\, ,\nn
\end{eqnarray}
where
\begin{equation}
{\cal D}_{\mu} \epsilon= D_{\mu} \epsilon-\frac{3i}{4}\gamma_5 A_\mu\eps \, .
\end{equation}
These are precisely the transformations of $d=4, \ N=1$ conformal
supergravity. This agrees well with the results of \cite{decoupling}
where it was found that the $AdS_5\times S^5$
superisometries reduce to superconformal transformations on the
boundary of the $AdS$ space.

\vskip 0.15in
\paragraph{$\bullet$ Summary}
We have shown that the gravity multiplet and symmetries of
four-dimensional, $N=1$ conformal supergravity  are induced on
the boundary of solutions to pure $N=2$ gauged supergravity in five
dimensions.  Similarly, $N=2k$ gauged supergravity in five dimensions
can be related to 4-dimensional $N=k$ conformal supergravity.  It is
worth asking whether the induced action on the spacetime boundary
respects the conformal gravity symmetries.  This action is
generally divergent and requires regulation. The regulator may be
chosen to preserve Weyl invariance, yielding an induced conformal
theory of gravity.  We will argue that if Weyl invariance
is explicitly broken by, say, putting in a radial cutoff as in the
Randall-Sundrum scenario, the action of 4d Poincar\'e gravity can be
induced on the boundary.

\setcounter{equation}{0}
\section{Conformal Yang-Mills and Conformal Gravity}
\label{cym}

In order to study the action induced on the spacetime boundary, it is
convenient to work within the AdS/CFT correspondence with relates 4d,
$N=4$ conformal gravity and 5d, $N=8$ gauged
supergravity~\cite{juanads}.  The conventional Lagrangian for
Yang-Mills on a curved manifold ($L= \sqrt{g} g^{mk} g^{nl} F_{mn}
F_{kl}$) enjoys a generalization with local $\CN = 4$ superconformal
invariance (see the review~\cite{csgrev} and the recent work of Liu
and Tseytlin~\cite{liutsey}).\footnote{This section begins by
summarizing standard results regarding conformal Yang-Mills coupled to
conformal gravity, as transmitted to recent audiences by Liu and
Tseytlin~\cite{liutsey}.}  For a single $SU(N)$ vector multiplet
$(A_m, \psi_i, X_{ij})$\footnote{$i,j = 1,2,3,4$ are indices of the
SU(4) R-symmetry.} the Lagrangian is
\begin{eqnarray}
L_{{\rm SYM}} =
-{1 \over 4} (e^{-\phi} F^{mn} F_{mn} + \CC F^{mn} F_{mn}^* ) -
{1 \over 2} \bar{\psi}^i \gamma^m D_m\psi_i -
{1 \over 4} X_{ij}(-D^2 + {1\over 6} R) X^{ij} \\
- X_{ij} F^{+mn} T^{ij}_{mn} +
\CD_{ij}^{kl} (X^{ij} X_{kl} - {1\over 6} \delta_k^i \delta_l^j |X|^2)
+ \cdots +
{\rm h.c.}
\label{YMback}
\end{eqnarray}
The coupling constants $g_{mn}$, $\varphi = e^{-\phi} + i C$, etc.,
are in superconformal representations that fill out the field content
of $\CN =4$ conformal supergravity, but appear here as constant
backgrounds rather than dynamical fields.

The effective action for the Euclidean field theory as a function of
coupling constants is computed by integrating out the Yang-Mills
fields, and gives a divergent part and a finite part:
\begin{equation}
W(g_{mn},\varphi,\cdots) =
\int \CD A \, \CD\psi \, \CD X
e^{-S_{SYM}}
\equiv W_{{\rm div}} + W_{{\rm fin}}\, .
\end{equation}
The divergences that arise despite the conformal invariance of the
theory are related to contact singularities in the definition of
composite operators.  Power law divergences can be cancelled by
local counter-terms, but the effective action will contain a
logarithmic divergence, $W_{{\rm div}}$, associated with the four
dimensional conformal anomaly.\footnote{There are potential
quartic and quadratic divergences which are proportional to an
effective cosmological constant and the Ricci scalar of the
manifold.} Let us regulate this divergence by introducing a
spatially uniform, covariant, Euclidean point-splitting cutoff:
the endpoints of propagators in a Feynman diagram must be
separated by a geodesic length exceeding some $\epsilon$. When
$\epsilon$ is very small this can be written:
\begin{equation}
g_{mn}(\bx) \, \Delta x^m \, \Delta x^n \geq \epsilon^2 \, .
\label{cutoff1}
\end{equation}
$W_{{\rm div}}$ diverges in the limit that $\epsilon$ vanishes.   The
difference in actions computed for $\nu$ vector multiplets, and with
cutoffs $\epsilon$ and  $\bar\epsilon$ is:\footnote{See~\cite{liutsey}
and
references therein.}
\begin{eqnarray}
W_{div} &=& {\nu \over 4(4\pi)^2} \,
\ln\left({\epsilon \over \bar{\epsilon}}\right) \,
\int d^4x \,
\sqrt{g} \, L_{{\rm CSG}} \label{divterm}\\
L_{{\rm CSG}} &=& C_{mnkl} C^{mnkl} - E +
4[D^2\varphi^* D^2\varphi - 2(R^{mn} - {1\over 3} g^{mn} R)
D_m\varphi^* D_n\varphi] +  \cdots
\label{CSGlag}
\end{eqnarray}
Here $C$ is the Weyl tensor, $E$ is the Euler invariant, and $C^2 - E
= 2(R_{mn} R^{mn} - R^2/3)$. $L_{{\rm CSG}}$ is precisely the
Lagrangian for four dimensional $N=4$ conformal supergravity.  (We will take
$\nu = N^2 - 1 \approx N^2$ for an SU(N) Yang-Mills theory at large
$N$.)  Integrating out the Yang-Mills fields has ``induced'' a Weyl
invariant theory of gravity on the manifold.

Classically, the Lagrangian (\ref{YMback}) is conformally invariant
and is independent of the Weyl factor in the background metric --
equivalently,  the trace of the classical stress tensor
vanishes.  However, the logarithmic divergence in (\ref{CSGlag}) results
in an anomalous dependence on the Weyl factor of the metric and
results in an anomalous trace in the stress tensor:
\begin{equation}
T = {\nu \over 2 (4\pi)^2} (R_{mn} R^{mn} - {1 \over 3} R^2 + \cdots)
\label{anomtr}
\end{equation}
The ellipses denote terms that appear when the other couplings in
(\ref{YMback}) such as $\phi$, $C$ etc. are spatially varying.  In
fact, the right hand side of (\ref{anomtr}) is proportional to the
conformal supergravity Lagrangian (\ref{CSGlag}).

Accordingly, the finite part of the effective action $W$ will contain
an anomalous piece that depends on the Weyl factor of the metric and
whose variation produces the trace (\ref{anomtr}).  Including the Weyl
invariant piece $W_{{\rm inv}}$ gives the finite part of the action:
\begin{equation}
W_{{\rm fin}} = W_{{\rm anom}} + W_{{\rm inv}}
\label{finterms}
\end{equation}
The Weyl invariant piece will be a series in even powers of $\epsilon$
because the curvature invariants forming the metric-dependent part of
the action have dimension two:
\begin{equation}
W_{{\rm inv}} = W_0 + \epsilon^2 W_2 + \epsilon^4 W_4 + \cdots
\label{finexp}
\end{equation}
The higher order terms  vanish as the cutoff is removed and
are regularization scheme dependent.

The conformal anomaly on the other hand is essentially scheme-independent,
barring a term proportional to $\nabla^2 R$ in (\ref{anomtr}) whose
regularization-dependent coefficient we have set to zero.  Having fixed
this ambiguity, $W_{{\rm anom}}$ has a diffeomorphism invariant, but
non-local, expression whose Weyl variation produces the anomaly
(\ref{anomtr}) (see \cite{riegert}, \cite{deser,deform}).  However,
splitting the metric into a Weyl factor and a reference background,
\begin{equation}
g_{ij} = e^{2\sigma} \, \bar{g}_{ij} \, ,
\end{equation}
yields a local expression for the dependence of $W_{{\rm anom}}$ on
$\sigma$~\cite{riegert, liutsey}\footnote{Note that the standard Riegert 
action employed here is potentially problematic in dimensions $d>2$ 
\cite{deser}.}:  
\begin{eqnarray}
W_{anom} = -{\nu \over 2(4\pi)^2} \,
\int d^4x \, \sqrt{\bar{g}} \, \left[
(\bar{R}^2_{mn} - {1 \over 3} \bar{R}^2  + 2 \bar{D}^2\varphi^*
\bar{D}^2\varphi + \cdots ) \sigma \right. +
\label{riegloc1} \nn \\
\left.
+ 2 \bar{G}^{mn} \bar{D}_m\sigma \, \bar{D}_n\sigma +
2 \bar{D}^{m}\sigma \, \bar{D}_m \sigma \, \bar{D}^2\sigma +
(\bar{D}^m \sigma \, \bar{D}_m \sigma)^2 \right]
 \label{riegloc2}
\end{eqnarray}
Here $\bar{G}_{mn}$ is the Einstein tensor of $\bar{g}_{mn}$ and
the terms linear in $\sigma$ in (\ref{riegloc1}) are precisely the
conformal supergravity action of the reference metric
$\bar{g}_{ij}$. The trace of the stress tensor can be expressed
in terms of $\sigma$ and $\bar{g}_{mn}$ as $T = -
(e^{-4\sigma}/\sqrt{\bar{g}}) (\delta W/\delta \sigma)|_{\sigma =
0}$ which reproduces (\ref{anomtr}). In fact, the divergent part
of the effective action and the quadratic part of anomalous piece
can be conveniently combined into a single, closed-form,
non-local action.\footnote{See \cite{riegert, deser, liutsey, deform} and
references therein.}

The logarithmic divergence and the related finite anomalous term are
exact at 1-loop for $\CN = 4$ Yang-Mills (see the references
in~\cite{liutsey}) and so we can reliably extrapolate the weak
coupling results above to the large 't Hooft coupling, large $N$
Yang-Mills theory which should be related to a classical gauged
supergravity.  According to the AdS/CFT correspondence, $W$, the
effective action of the Yang-Mills theory as a functional of sources,
is equal to the classical action of the bulk gauged supergravity as a
functional of boundary data.  So we have just shown that to leading
(logarithmic) order the 5d $N=8$ supergravity induces $N=4$ conformal
supergravity on the spacetime boundary.  However, as we have seen, the
effective action also contains finite anomalous and Weyl invariant
terms.    According to AdS/CFT, these terms should together reconstruct
the dynamics of five dimensional $N=8$ gauged supergravity.

Unfortunately, although parts of the finite, Weyl invariant part of the
action ($W_{{\rm fin}}$ in (\ref{finterms})) are fixed by the classic
analyses of the conformal anomaly, $W_{{\rm fin}}$ also contains terms that
are not under control in the strong coupling limit.  For example, the power
series in $\epsilon$ in (\ref{finexp}) could be greatly modified. So we
will try to milk the anomaly for as much data as possible in reconstructing
the bulk spacetime from field theory.

\subsection{The Anomaly and the Renormalization Group}

The renormalization group studies the transformation of the effective
action $W$ as a function of the cutoff $\epsilon$.  The basic idea is
that a change of the cutoff $\epsilon \rightarrow \epsilon'$ is
equivalent to a redefinition of the couplings $g \rightarrow g'$ at
fixed cutoff.  The resulting effective variation of the couplings as a
function of the cutoff is described by the RG equation.

For example, we will show that redefining the cutoff $\epsilon$ in
(\ref{cutoff1}) by a spatially varying factor $e^{\lambda(x)}$ is
equivalent, to leading order in $\epsilon$, to re-scaling the
metric by a Weyl factor:
\begin{eqnarray}
g_{mn} \, \Delta x^m \, \Delta x^n \geq
\epsilon^2 \, e^{2\lambda(x)} ~~~~~ \Longrightarrow ~~~~~
\tilde{g}_{mn} \, \Delta x^m \, \Delta x^n \geq \epsilon^2
\label{tradeoff}
\\
\tilde{g}_{mn} \equiv
e^{-2\lambda(x)} \, \left[ g_{mn} + \nabla_m V_n(\bx,\epsilon) +
\nabla_n V_m(\bx,\epsilon) \right]
\label{newmet}
\end{eqnarray}
Here we have permitted general $\epsilon$-dependent diffeomorphisms of
the manifold generated by the vector field $V^i(\bx,\epsilon)$ since
these are symmetries of the theory.

When $\lambda$ is constant it is easy to show the equivalence in
(\ref{tradeoff}) for the leading terms in the effective action --
namely, the anomalous and logarithmic pieces.  Re-scaling $\epsilon$
by $e^\lambda$ shifts the log-divergent term (\ref{divterm}) by
$\lambda \nu / 4(4\pi)^2$ times the conformal supergravity action.
The same shift is produced in the $W_{{\rm anom}}$ by re-scaling the
metric by $e^{-2\lambda}$.  However, $W_{{\rm div}}$ is left invariant
by a Weyl re-scaling of the metric so long as the other fields in the
conformal gravity Lagrangian (\ref{CSGlag}) are also re-scaled by
their Weyl weights.  For example, the scalar $\varphi$ has weight zero and
therefore remains invariant.  So, when $\lambda$ is constant, a change
in the cutoff can be traded, in the leading terms of the effective
action, for a Weyl re-scaling of the metric and a corresponding
re-scaling of all the couplings in (\ref{YMback}) by their Weyl
weights.  However, the cutoff dependence of the Weyl invariant finite
terms (\ref{finexp}) implies that keeping the entire action invariant
will require more than a Weyl transformation of the fields -- in the
perturbative limit, the couplings have to be corrected at each order
in $\epsilon$ to keep the entire effective action invariant.  These
higher order corrections cannot be reliably extrapolated to the strong
coupling limit, but non-renormalization of the conformal anomaly
guarantees that as $\epsilon \rightarrow 0$, the equivalence
(\ref{tradeoff}) is valid.

It is a little harder to argue that this is still the case for a spatially
varying cutoff, because the action (\ref{divterm}) is explicitly computed
for a constant cutoff.  Instead, examine the origin of divergences and the
anomaly in logarithmic singularities that occur when two points in a
Feynman diagram approach each other closely.  The cutoff in
(\ref{tradeoff}) restricts the proximity of such points by placing a lower
bound on the size of vectors $\Delta X^m$ in the tangent space at $\bx$.
Re-scaling the cutoff increases the bound on $\Delta X^m$.  Since the
classical Lagrangian in (\ref{YMback}) is both Weyl and diffeomorphism
invariant, the Feynman diagrams are not changed by a combined re-scaling of
the metric as in (\ref{newmet}) and the other couplings by the appropriate
Weyl weights.  The only effect of this redefinition of couplings with a
fixed small cutoff $\epsilon$ is to re-scale the bound on the size of
vectors $\Delta X^m$ measuring separation between nearby points in the
Feynman diagram integrations.  In other words, at the level of the
diagrammatic computation of the log divergent part of the effective action,
a small spatially varying cutoff can be directly traded for a Weyl-rescaled
metric as in (\ref{tradeoff}).  The anomalous terms in the action can be
deduced from this following~\cite{riegert}.  In this argument, it is
essential that we understand the cutoff $\epsilon$ to be both small and
slowly varying -- (\ref{tradeoff}) is a covariant cutoff only under these
conditions.

We have just argued that that a small, slowly varying cutoff can be traded for a
redefined metric in a way that leaves the sum of the
log-divergent and anomalous terms in the action invariant.
The other couplings appearing in the effective action are also
re-scaled according to their Weyl weights; e.g., the scalar $\varphi$
has weight $0$ and remains invariant.

In fact,  knowing the trace anomaly (\ref{anomtr}) of the theory 
fixes the anomalous part of the action and the log divergence up to
Weyl invariant terms.  First, it is possible to integrate the
trace anomaly to find a diffeomorphism invariant action that
varies to the anomaly \cite{riegert,deser,deform} (also see the
references in~\cite{liutsey}).  This action is not unique -- any
local or non-local Weyl invariant may be added to it without
changing the trace anomaly~\cite{deser,deform} and other methods are
required to determine these terms.  Then, reversing the logic
above we can infer the presence of a divergence logarithmic in
the cutoff.

In essence, integrating out the Yang-Mills theory has ``induced'' a
Weyl-invariant gravitational action on the manifold.  According to the
AdS/CFT correspondence, and in parallel with Sec.~2, this action must also
be induced by the bulk $N=8$ gauged supergravity.  Below we will see how
far we can go towards showing this directly from the bulk perspective.

\subsection{Gravity description}
Happily, the on shell massless fields of five dimensional $\CN = 8$ gauged
supergravity have precisely the multiplicities of the couplings in
(\ref{YMback}), and transform in the same way under the asymptotic
(super-conformal) symmetry group of the gravitational
theory~\cite{csgrev,group}.  Following the AdS/CFT prescription we should
compare the classical action for the 5-dimensional supergravity as a
functional of boundary data to the Yang-Mills effective action $W$.  

The cutoff length scale $\epsilon$ that appears in the field theory
effective action is related to radial positions in the bulk
space~\cite{susswitt, amanda,adsrg,adsrgothers,hermanrg}.  Indeed, when
$\epsilon$ is small and slowly varying as a function of boundary positions
$x$, it can be directly identified with a radial
cutoff~\cite{adsrg,hermanrg,deform}.  The field theory scheme
dependence of choosing spatially varying cutoffs $\epsilon(x)$ is directly
related to truncations of  the bulk space by ``wavy'' surfaces
parameterized as $r(x) = \epsilon(x)$ in the coordinates (\ref{gensoln}).
All of these surfaces are related by five dimensional diffeomorphisms, and
the metric induced on them is given precisely by
(\ref{newmet})~\cite{deform}.  In other words, diffeomorphisms of the 5d
spacetime are directly related to a choice of RG scheme for the dual field
theory!

Henningson and Skenderis~\cite{HenSken} showed that the gravitational
terms in the 5d action contain quartic and quadratic divergences and a
logarithmic divergence equal to the gravitational terms in
(\ref{divterm}).  It was shown in~\cite{HenSken, adsstress} that the
power law divergences could be cancelled by local boundary
counterterms in the gravitational action.  The leading piece of the
bulk action as a functional of boundary data is then the
logarithmically divergent term equal to the gravitational part of the
four dimensional conformal anomaly\footnote{In \cite{Nojiri} this analysis
was extended to dilatonic gravity.}. 
The results of~\cite{deform}
imply that this analysis continues to hold for a general foliation of
the bulk spacetime by ``wavy'' cutoff surfaces.  In other words, the
leading gravitational terms in the five dimensional action exactly
reproduce a conformally invariant action for boundary gravity as
implied by the Yang-Mills conformal anomaly.  In fact, these leading
terms arise from the action accumulated by the divergent behaviour of
the metric near the boundary of the bulk space; finite energy
excitations contribute subleading terms because, as implied by the
results of~\cite{adsstress}, they contribute to finite parts of the
action.  So we learn that the complete gravitational dynamics of the
skin of an asymptotically $\ads{5}$ space is contained the in the
four-dimensional conformal anomaly.

The above discussion was carried out purely for the gravitational
terms in the bulk and boundary actions.  However, it is expected that
inclusion of the scalars, fermions and gauge fields of 5d $N=8$
supergravity would induce the full $N=4$ conformal supergravity action
on the boundary of the space.  In the previous section we showed that
5d, $N=2$ gauged supergravity induces the symmetries of $N=1$
superconformal gravity on the boundary.  Here we expect (although it
is technically much harder to show) that the symmetries of $N=4$
conformal gravity are induced on the boundary.  Given these
symmetries, transformations of the gravitational terms ($C^2 - E$) are
expected to give the remaining terms of the $N=4$ conformal gravity
Lagrangian.

\vskip 0.15in
\paragraph{$\bullet$ Summary}
We have used the AdS/CFT correspondence to argue that $N=4$ conformal
gravity is induced on the 4-dimensional boundary of solutions to
5-dimensional $N=8$ gauged supergravity.  Turning things around, we have also
argued that the 4d conformal anomaly encapsulates the gravitational
dynamics of the skin of asymptotically $\ads{5}$ spaces.  Conformal
gravities also exist in odd dimensions where there is no conformal anomaly.
In these situations which arise, for example, in the $\ads{4}/{\rm CFT}_3$
correspondence, the bulk action does not have a log divergence and must
induce a finite conformally invariant action on the boundary.

\setcounter{equation}{0}
\section{Discussion - The Role of Conformal Gravity}
\label{discussion}

    To summarize, we have demonstrated, with an explicit mapping of
symmetry actions, how $N=2$ gauged supergravity in the bulk of AdS
space induces $N=1$ conformal supergravity on a boundary surface.  The
explicit breaking of Weyl symmetry involved in restricting to a
cut-off surface will only add small perturbations to this scheme if we
keep the surface near the boundary of AdS.  With additional algebraic
complexity a similar induction of $N=4$ boundary conformal
supergravity should follow from bulk $N=8$ gauged supergravity. This
is also well motivated from our discussion of the anomaly structure of
$N=4$ SYM.  Below we will argue that when we use boundaries far inside
AdS, Weyl-symmetry is strongly broken and the boundary theory is no
longer conformal.  In this situation, which is exploited in the
Randall-Sundrum model, the induced gravity is only Poincar\'e
invariant.

\vskip 0.15in
\paragraph{$\bullet$ Supersymmetric Counterterms and Holographic RG Flows}
Our results regarding $N=2$ gauged supergravity have notable
applications to the derivation of supersymmetric boundary counterterms
for AdS gravity and holographic RG-flows derived from the resulting
regulated actions.

It has been shown that the power law divergences in the action and
stress tensor of a space that is asymptotically locally AdS can be
eliminated by the introduction of intrinsic boundary
counterterms~\cite{adsstress}.  These methods avoid various
ambiguities and technical difficulties associated with other methods
in classical gravity for computing the action and conserved charges of
a space.  Using the induced $N=1$ SUSY boundary transformations that we have
derived, one could compute the counterterms for the entire gravity
supermultiplet by transforming the gravitational counterterms
of~\cite{adsstress}.

Also, the AdS/CFT correspondence states that the classical action for
the bulk space, regulated by these boundary counterterms and seen as a
functional of boundary data, is equal to the effective action for an
$N=1$ SYM theory that is conformal in the ultraviolet.  This theory
has a conformal anomaly which, in addition to the familiar Weyl tensor
squared and Euler invariant contributions ($C^2 - E$), includes terms
involving scalars and fermions.  By the AdS/CFT correspondence,
this must be equal to a log divergent term of the 5d spacetime action.
Acting on $C^2 - E$ with the explicit $N=1$ superconformal
transformations that we have identified will yield the complete strong
coupling supersymmetric conformal anomaly of the dual $N=1$ SYM
theory.

In Sec.~3.1 we discussed the matching between field theory cutoffs and
radial positions in the bulk space.  This is the basis of the
holographic renormalization group~\cite{adsrg,adsrgothers,hermanrg}.  By
matching the bulk $N=2$ SUSY with the boundary $N=1$ SUSY for the
theory at each length-scale of a given renormalization flow, our
methods can also provide useful tools for the holographic analysis of
Shifman-Vainshtein relations between supersymmetric beta functions.

\vskip 0.15in
\paragraph{$\bullet$ Induced Poincar\'e Gravity -- the $N=1$ Supersymmetric
Randall-Sundrum Model}
Thus far we have discussed how Weyl-invariant gravity is induced on
the boundary of spaces governed by gauged supergravity.  Our
considerations are also relevant to analyses of Randall-Sundrum
models where the Standard Model is attached to a domain wall in five
dimensional AdS space on which four dimensional Poincar\'e invariant
gravity has been localized~\cite{rs}.

We showed in Sec.~\ref{onesusy} that radial diffeomorphisms of five
dimensional gauged supergravity induce Weyl transformations of the
surfaces homeomorphic to the spacetime boundary.  If Weyl symmetry is
maintained as a residual symmetry on such surfaces, the induced
gravitational action is conformally invariant (up to an anomaly).
In other words, in the computation of the on-shell bulk
supergravity action the counterterms~\cite{adsstress}, which make the
bulk action finite, also precisely cancel induced boundary quantities that
break the residual Weyl symmetry such as the Einstein-Hilbert and
cosmological terms.  Weyl invariance is nevertheless broken
anomalously as in Sec.~\ref{cym} by the radial cutoff dependence of
a logarithmic divergence which cannot be cancelled.

However, in situations like the Randall-Sundrum scenario~\cite{rs}, Weyl
invariance is explicitly broken by the choice of a fixed radial position in
AdS space where a brane is placed.  In this case, there is no reason to
pick a regulation scheme such as~\cite{adsstress} which preserves the
residual Weyl symmetry.  Indeed, the Einstein-Hilbert counterterm
in~\cite{adsstress} can be ignored completely, allowing the bulk to induce
Poincar\'e invariant gravity on fixed-radius surface.  Then, the results of
Sec.~\ref{onesusy} can be used to study the $N=1$ supersymmetric structure
induced by the bulk theory. However, the residual $N=1$ supersymmetry still
forbids the appearance of a cosmological constant.  This reasoning also
applies to extended supersymmetry. An $N=2k$ supersymmetric bulk
supergravity extension of the Randall-Sundrum model will ``induce'' an
$N=k$ Poincar\'{e} supergravity on the wall (see also~\cite{dennis}).

\vskip 0.15in
\paragraph{$\bullet$ Dynamical conformal gravity?}
In this paper we have discussed the appearance of conformal gravity at
the boundary of spaces governed by gauged supergravity.  The fields
that appeared in this discussion were the boundary values of
non-normalizable bulk modes that appear in asymptotically AdS
spaces~\cite{bkl}.  These modes cannot fluctuate unless the space is
truncated at a finite radius, because their action is infinite if the bulk
is non-compact.  Therefore, the conformal gravity induced on the
boundary of AdS space is not dynamical.  Equivalently, as discussed in
Sec.~3, conformal gravity appears from the AdS/CFT perspective as the
effective action of a field theory and is a functional of sources.
Such effective actions do not describe dynamical theories -- they are
merely generating functions for correlators and should not be varied
to compute equations of motion.

However, if the bulk space is cut off at a finite distance, the
(formerly) non-normalizable modes that we have used to induce
conformal gravity will have finite actions.  Therefore they will be
able to fluctuate. This suggests that they should become actual
dynamical fields on the cut-off surface, and that dynamical conformal
gravity is the effective theory observed by an experimentalist placed
on a surface of fixed radius near the boundary of an asymptotically
AdS space\footnote{Dynamical gravity on the brane was also
considered in \cite{Abou-Zeid}.}.  
A sigma model of closed strings in AdS space presumably
includes worldsheets with boundaries attached to the spacetime
boundary.  Such string configurations are the natural sigma model
analogues of the non-normalizable modes.  It would be natural to
expect that they are responsible for inducing dynamical conformal
gravity on surfaces near an AdS boundary.

The matter is subtler from the perspective of the dual CFT.  In the
large $N$ limit, we usually equate the SYM effective action with a
path integral over bulk supergravity fields subject to boundary
conditions on the $AdS$ boundary.  These boundary conditions are
implemented by choosing a non-normalizable mode background.
Regulating the SYM theory is equivalent to cutting off the $AdS$ space
at some finite distance. At this cut-off surface, the values of
supergravity fields are cannot be fully fixed and should be integrated
over.  Since boundary values of the supergravity modes correspond to
sources in the SYM~\cite{bkl}, it would seem that in the
regulated SYM effective action we must integrate over both the fields
and the sources, including the superconformal gravity multiplet. The
resulting path integral is not a functional of the sources anymore,
but rather a functional of the initial and final states. Hence, it
would compute an S-matrix for the modified CFT.  As the cutoff
$\epsilon \rightarrow 0$ in this modified path integral, we require
that the sources become frozen. This suggests an interesting
perspective: the cutoff $\epsilon$ in some sense translates to an
effective $\hbar$ for field theory sources within the AdS/CFT
correspondence.

\vskip.2in \centerline{\bf Acknowledgments}
\medskip

We have enjoyed discussions with R.~Corrado, B.~de Wit, J.~de Boer,
P.~Ho\v{r}ava, H.~Ooguri, K.~Pilch, L.~Randall, S.~Rey, K.~Sfetsos, K.~Stelle,
E.~Witten and especially A.~Tseytlin.  {\small V.B.}  was supported by
the Harvard Society of Fellows, the Milton Fund of Harvard University
and by NSF grant NSF-PHY-9802709.  {\small E.G.} and {J.R} were
supported by DOE grant DE-FG03-92ER40701 and the Caltech Discovery Fund. 
{\small D.M} was supported
by DOE grant DE-FG03-84ER40168, and thanks the University of Illinois
at Chicago for hospitality during the final part of this project.


\end{document}